\documentstyle[prl,aps,epsf]{revtex}
\begin{document}

\advance\textheight by 0.2in
\draft
\twocolumn[\hsize\textwidth\columnwidth\hsize\csname@twocolumnfalse
\endcsname  

\title{Critical Behavior of the Meissner Transition in the
Lattice London Superconductor}

\author{Peter Olsson$^1$ and S. Teitel$^2$ }
\address{$^1$Department of Theoretical Physics, Ume{\aa} University, 
901 87 Ume\aa, Sweden}
\address{$^2$Department of Physics and Astronomy, University of Rochester, 
Rochester, New York 14627}
\date{\today}
\maketitle

\begin{abstract}
We carry out Monte Carlo simulations of the three dimensional
(3D) lattice London
superconductor in zero applied magnetic field, making
a detailed finite size scaling analysis of the Meissner 
transition.  We find that the magnetic penetration
length $\lambda$, and the correlation length $\xi$, scale as
$\lambda\sim\xi\sim |t|^{-\nu}$, with $\nu=0.66\pm 0.03$,
consistent with ordinary 3D XY universality, $\nu_{\rm XY}
\approx 2/3$.  Our results confirm the anomalous scaling dimension of
magnetic field correlations at $T_c$.
\end{abstract}

\pacs{74.40.+k, 64.60.Fr}

]

The discovery of the high temperature superconductors has 
revived interest in the effects of fluctuations on
the critical behavior of the superconducting transition.
The Meissner transition of a bulk type II superconductor
in zero applied magnetic field is the most basic case that
can be considered.  While it was originally thought that
this transition was weakly first order \cite{R00}, it is 
now generally believed to be in the
same universality class as the ordinary three dimensional (3D)
XY model, except with the temperature scale inverted 
\cite{R0,R4,R2}.
The argument is based \cite{R0,R4,R2,R1,R3} on two observations: 
(i) the ordinary 3D XY
model can be mapped onto a system of sterically interacting loops
with inverted temperature scale $T_{\rm loop}\propto 1/T$, 
and (ii) the vortex loops of a fluctuating superconductor interact
with a $screened$ Coulomb interaction, with screening length equal
to the bare magnetic penetration length $\lambda_0$.  Assuming that
the finite interaction length $\lambda_0$ of the vortex loops is not
a relevant modification of steric (on site) interactions, the
universality of the fluctuating Meissner transition and the 3D XY 
model follows \cite{R5.5}.  Early Monte Carlo (MC) simulations by Dasgupta
and Halperin \cite{R4} of a lattice London
superconductor model strongly supported this picture by making a 
qualitative comparison of the shape of specific heat peaks in the two 
models. 

Recently there has been renewed interest in, and controversy 
concerning, the nature of this transition.  Kiometzis {\it et al.}
\cite{R5}, considering a dual formulation of the fluctuating 
Ginzburg-Landau (GL) model, have argued that while the correlation length 
$\xi\sim |t|^{-\nu}$
diverges with the same exponent $\nu$ as the ordinary 3D XY model,
$\nu_{\rm XY}\approx 2/3$, the renormalized magnetic penetration length
should diverge as $\lambda\sim |t|^{-\nu^\prime}$ with $\nu^\prime=1/2$
the mean field exponent.  Herbut and Te\v{s}anovi\'{c}  \cite{R6} however,
using an analysis of the GL model exact to all orders in perturbation, have
argued that, due to the presence of an {\it anomalous
dimension}, $\eta_A=1$, for fluctuations of the magnetic field, one must
have $\nu=\nu^\prime$.  Using a one-loop renormalization group (RG)
scheme, they further suggested the possibility that $\nu <\nu_{\rm XY}$
\cite{R6.5}.  Bergerhoff {\it et al.} \cite{R8}, using a non-perturbative
RG flow analysis of the GL model, similarly find
$\nu<\nu_{\rm XY}$.  Herbut \cite{R7} however has argued that for
the lattice London limit of the GL model, $\nu=\nu^\prime=\nu_{\rm XY}$.

To investigate this controversy we present here the results
of new MC simulations of the 3D isotropic lattice London 
superconductor (LLS) in zero external magnetic field.
Carrying out the first detailed finite size scaling
analysis of this model, we find results 
consistent with a $single$ diverging
length scale, hence $\xi\sim\lambda$.  We find $\nu\approx\nu_{\rm 
XY}$ consistent with the universality of the ordinary
3D XY model.  We find clear evidence for the anomalous dimension
of magnetic field correlations predicted by Herbut and Te\v{s}anovi\'{c}
\cite{R6}.

The Hamiltonian of our model \cite{R4} is 
\begin{equation}
  {\cal H}=\sum_{i\mu}\left\{
  U(\theta_{i+\hat\mu}-\theta_i-A_{i\mu})
  +{1\over 2}J\lambda_0^2[{\bf D}\times{\bf A}]_{i\mu}^2\right\}
  \enspace .
\label{eH}
\end{equation}
The sum is over all bonds of a 3D simple cubic lattice of unit
grid spacing.  $\theta_i$ is the phase angle of the superconducting
wavefunction on site $i$, $\psi_i=e^{i\theta_i}$, where the
amplitude of $\psi_i$ has been taken constant (the London
approximation).  $A_{i\mu}$ is the discretized vector potential 
on the bond at site $i$ in direction $\hat\mu=\hat x,\hat y, \hat z$, 
and if $\mu,\nu,\sigma$ is a cyclic permutation of $x,y,z$, then
\begin{equation}
   [{\bf D}\times{\bf A}]_{i\mu}
   = A_{i\nu}+A_{i+\hat\nu,\sigma}-A_{i+\hat\sigma,\nu}-A_{i\sigma}
   \equiv 2\pi b_{i\mu}
\label{eDtA}
\end{equation}
is the counterclockwise circulation of the $A_{i\mu}$ around the 
plaquette at site
$i$ with normal in direction $\hat\mu$.
$b_{i\mu}$ is the number of flux quanta $\phi_0$ of total
magnetic field through this plaquette.  The coupling is 
$J=\phi_0^2/16\pi^3\lambda_0^2$, with $\lambda_0$ the {\it bare}
magnetic penetration length, and $U(\varphi)$ 
is the Villain function \cite{R9}
$$e^{-U(\varphi)/T}=\sum_{m=-\infty}^\infty 
e^{-{1\over 2}J(\varphi-2\pi m)^2/T}\enspace.$$
The first term in Eq.\,(\ref{eH}) is the kinetic energy of 
flowing supercurrents; the second term is the magnetic field
energy.

We focus here on the calculation of the magnetic field correlation
function
\begin{equation}
  F(q)\equiv {4\pi^2J\over TL^3}\langle 
  b_\mu(q\hat\nu)b_\mu(-q\hat\nu)\rangle\enspace,
\label{eF}
\end{equation}
where $b_\mu(q\hat\nu)\equiv\sum_ie^{-iq\hat\nu\cdot{\bf r}_i}b_{i\mu}$
is the Fourier transform of the total magnetic field, and $\hat\mu
\perp\hat\nu$.  $F(q)$ is just the wavevector dependent magnetic 
permeability, with $\lim_{q\to 0}F(q)=\partial B/\partial H$ \cite{R10}.
Our goal is to show that the singular part of $F(q)$ is consistent 
with the scaling Ansatz
\begin{equation}
  F(t,Q,L)=\ell^{-1}F(t\ell^{1/\nu},Q\ell,L/\ell)\enspace,
\label{eFscale}
\end{equation}
where $t=T-T_c$, $Q=2\sin(q/2)$, $L$ is the system length, and 
$\ell$ is an arbitrary length rescaling factor. 
Note that we choose $Q$ rather than $q$ as our scaling
variable, since the vortex line interaction that arises
from the Hamiltonian (\ref{eH}) is a function of $q_\mu$
only through the combinations $Q_\mu$ \cite{R10}.  Since
$Q\to q$ as $q\to 0$, this does not affect the long length
scaling; our hope is that by using $Q$ we may succeed to
slightly extend the scaling region to shorter length scales.
Verification of the scaling Eq.\,(\ref{eFscale}) will
demonstrate that there is only a $single$ diverging length scale in
the model, that describes both the critical behavior of global
thermodynamic variables, as well as the spatial variation of
magnetic field fluctuations.  Since the former is determined by
the correlation length $\xi$, while the later is determined by
the magnetic penetration length $\lambda$, we conclude that
$\xi\sim\lambda\sim |t|^{-\nu}$.

We carry out standard Metropolis MC on the Hamiltonian
(\ref{eH}) for cubic lattices of lengths $L=8$ to $32$, using
periodic boundary conditions.  We use the particular
value of $\lambda_0=0.3$ (in units of the grid spacing).  
Temperatures will be measured in units of $J$.  In one MC ``pass''
we first update $A_{ix}$, $A_{iy}$ and 
$A_{iz}$ at each site $i$, going sequentially through
the entire lattice, then follow this by a sequential
update of the $\theta_i$.  The
$A_{i\mu}$ are allowed to fluctuate without constraint.  
For our largest system size, $L=32$, we use at each temperature
typically $32,000$ passes to equilibrate, followed by 
$1.7\times 10^7$ passes for computing averages.

\begin{figure}
\epsfxsize=7.5truecm
\epsfbox{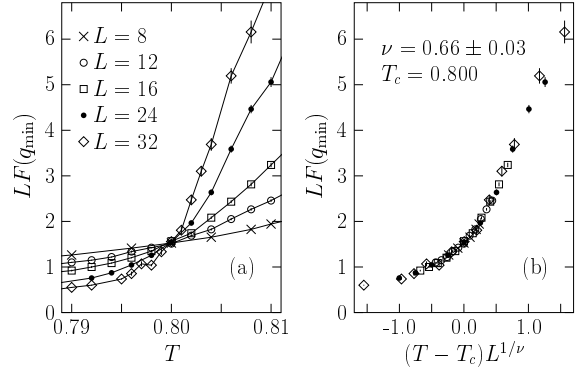}
\caption{a) $LF(q)$ vs.\ $T$ for system sizes $L=8-32$.
The common intersection of all curves locates $T_c$. 
b) Scaling collapse of $LF(q)$ vs.\ $(T-T_c)L^{1/\nu}$ using 
fitted values of $T_c$ and $\nu$.
}
\label{f1}
\end{figure}

First we consider the scaling behavior of the magnetic
permeability.  Evaluating Eq.\,(\ref{eFscale}) at the
smallest wavevector in our system, $q_{\rm min}=2\pi/L$, 
using $Q_{\rm min}\approx q_{\rm min}$, and choosing
the rescale factor $\ell = L$, we arrive at
\begin{equation}
  LF(t,Q_{\rm min},L)=F(tL^{1/\nu},2\pi,1)\enspace.
\label{eFqmin}
\end{equation}
Exactly at $T_c$ (i.e. $t=0)$, $LF(q_{\rm min})$ 
should thus be a constant independent of $L$.
In Fig.\,\ref{f1}a we plot our data for $LF(q_{\rm min})$ vs.\
$T$, for $L=8-32$.  To a very good accuracy, 
the curves for different $L$ do indeed intersect at a single point, 
$T_c\approx 0.8$.  To further verify the scaling relation 
Eq.\,(\ref{eFqmin}), we fit our data for $LF(q_{\rm min})$ near $T_c$
to a low order polynomial expansion in $(T-T_c)L^{1/\nu}$.
We determine the values of $T_c=0.8000\pm 0.0002$ and 
$\nu=0.66 \pm 0.03$ from a $5$th order polynomial fit, restricting
data to the ranges $|t|\le t_{\rm max}=0.006$ and $L\ge L_{\rm min}=12$.
Increasing either the order of the polynomial, $L_{\rm min}$, or decreasing
$t_{\rm max}$ resulted in no change in these fitted values, within
the estimated statistical error.
In Fig.\,\ref{f1}b
we use these fitted parameters to plot $LF(q_{\rm min})$
vs.\ $tL^{1/\nu}$, for data in the range $|t|\le 0.01$. 
The resulting data collapse is $very$ good.
Our value of $\nu$ is thus completely consistent with 
$\nu_{\rm XY}\approx 2/3$.  Note that by taking $q=q_{\rm min}$,
$L\to\infty$, and $\ell=t^{-\nu}$ in Eq.\,(\ref{eFscale}),
our results imply that the magnetic 
permeability vanishes, as $T\to T_c^+$, as $\partial B/
\partial H\sim t^\nu$.

\begin{figure}
\epsfxsize=7.5truecm
\epsfbox[92 322 586 660]{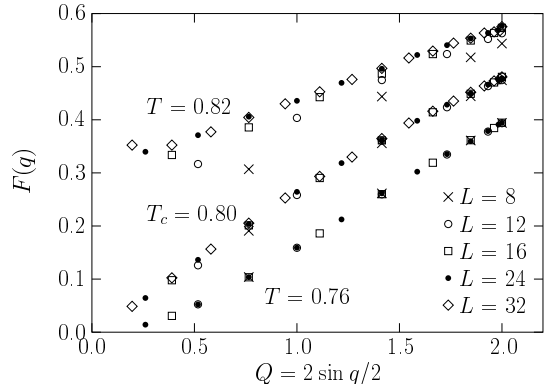}
\caption{$F(q)$ vs.\ $Q$ for system sizes $L=8-32$, at
$T=0.76<T_c$, $T_c=0.80$ and $T=0.82>T_c$.  
Note the virtual absence of finite size effects for
$T\le T_c$.
}
\label{f2}
\end{figure}

We now consider the $q$ dependence of $F(q)$.  
In Fig.\,\ref{f2} we plot $F(q)$ vs.\ $Q$ for 
$L=8-32$, exactly at $T_c$
and for one representative temperature above and below $T_c$.
We see, as expected, that for $T>T_c$, $F(q)$ approaches a 
constant as $Q\to 0$, while for $T<T_c$, $F(q)$ vanishes as $Q^2$.
Exactly at $T_c$ however, $F(q)$ appears to vanish {linearly} as $Q$.
This is a clear suggestion of the anomalous dimension of magnetic
field correlations predicted by Herbut and Te\v{s}anovi\'{c},
according to which at $T_c$, $F(q)\sim q^{\eta_A}$ with $\eta_A=4-D$ 
in $D$ dimensions \cite{R6}.
It is interesting to note that, while there is a considerable finite size 
effect for $T>T_c$, finite size effects at a fixed value of $Q$
appear negligible
for all $T\le T_c$.  

\begin{figure}
\epsfxsize=7.5truecm
\epsfbox[92 322 586 660]{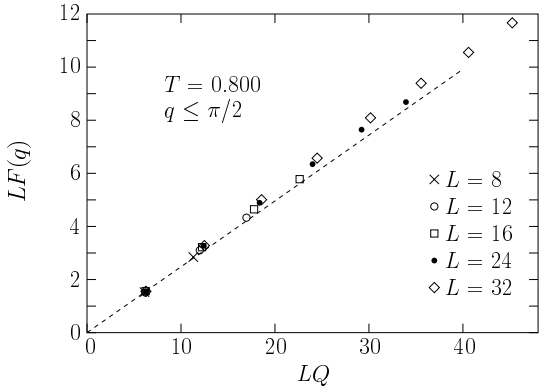}
\caption{Scaling collapse of $LF(q)$ vs.\ $LQ$ at $T_c$, for
$L=8-32$, $q<\pi/2$. $LF(q)$ vanishes $linearly$ as $LQ\to 0$.
}
\label{f3}
\end{figure}

To further verify the anomalous scaling dimension of magnetic
field correlations, we can apply Eq.\,(\ref{eFscale}) at $t=0$, 
taking as the rescaling factor $\ell=L$, to get
\begin{equation}
   LF(0,Q,L)=F(0,QL,1)\enspace.
\label{eFTc}
\end{equation}
In Fig.\,\ref{f3} we plot $LF(q)$ exactly at $T_c$ vs.\ $LQ$, for
$L=8-32$ and $q\le\pi/2$.  We find a good collapse
of the data to a single curve that vanishes linearly as
$LQ\to 0$.

\begin{figure}
\epsfxsize=7.5truecm
\epsfbox[92 322 586 660]{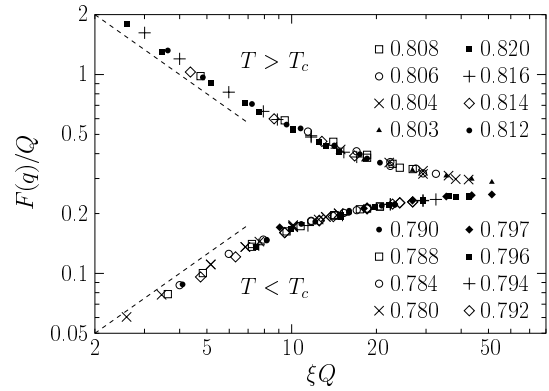}
\caption{Log-log scaling collapse of $F(q)/Q$ vs.\ $\xi Q$.  
The dashed lines at small $\xi Q$ have slopes of $\pm 1$,
to indicate the asymptotic behavior as $\xi Q\to 0$.
}
\label{f4}
\end{figure}

Finally, in the thermodynamic limit $L\to\infty$, we can use
Eq.\,(\ref{eFscale}) with $\ell =|t|^{-\nu}\equiv\xi$ to get
\begin{equation}
  F(t,Q,\infty)/Q=(Q\xi)^{-1}F_\pm(1,Q\xi,\infty)\enspace,
\label{eFxi}
\end{equation}
where $F_\pm$ refers to distinct branches for $T>T_c$ and $T<T_c$.
Using the values of $T_c$ and $\nu$ found in the fit of 
Fig.\,\ref{f1}b to determine $\xi=|T-T_c|^{-\nu}$, we plot in 
Fig.\,\ref{f4} our data for $F(q)/Q$ vs.\ $\xi Q$ on a log-log scale.  
We use only data for which finite size effects appear to be small, 
and which are in the scaling region.  
We see an excellent collapse of the data.  Fig.\,\ref{f4} clearly
demonstrates that there is only a single diverging length scale for
the spatial variation of magnetic field correlations, and that this
length scale is $\xi$.  For the $T>T_c$ branch, we see that $F(q)/Q$
diverges as $1/\xi Q$ as $\xi Q\to 0$, indicating that $F(q)$
approaches a finite constant $\propto\xi^{-1}$.  For the $T<T_c$ branch, we see
that $F(q)/Q$ vanishes as $\xi Q$ as $\xi Q\to 0$, indicating that
$F(q)$ vanishes as $\xi Q^2$.  However
for both branches
$F(q)/Q$ approaches the same constant as $\xi Q\to\infty$, indicating
that $F(q)$ vanishes linearly in $Q$ exactly
at $T_c$.  Fig.\,\ref{f4} thus gives another demonstration of the
anomalous dimension of magnetic field scaling at $T_c$.

To get a better physical understanding of the effects of
this anomalous dimension of magnetic field scaling, 
consider applying a small external magnetic
field given by ${\bf A}^{\rm ext}$.  The London equation,
describing the total screening of the Meissner state, $T<T_c$,
gives for the induced supercurrent \cite{R11}
\begin{equation}
  \langle j_{\mu}^{\rm ind}(q\hat\nu)\rangle ={J\lambda_0^2\over\alpha(q)}
  \langle A_{\mu}(q\hat\nu)\rangle\enspace,
\label{eLondon}
\end{equation}
where the vector potential of the total magnetic field
is the sum of the applied and induced fields, $\langle 
A_{\mu}(q\hat\nu)\rangle
=A_{\mu}^{\rm ext}(q\hat\nu)+\langle A_{\mu}^{\rm ind}(q\hat\nu)\rangle$, 
and 
\begin{equation}
\lambda_0^2/\alpha(q=0)=n_s(T)/n_s(T=0)
\label{ealpha}
\end{equation}
is determined by the density of superconducting electrons $n_s$.

The induced supercurrent is also related to $\langle A_{\mu}^{\rm ind}
(q\hat\nu)\rangle$
by Amp\`{e}re's Law, which for the gauge ${\bf Q}\cdot{\bf A}=0$
can be written as
\begin{equation}
  \langle j_{\mu}^{\rm ind}(q\hat\nu)\rangle=-J\lambda_0^2Q^2\langle
  A_\mu^{\rm ind}(q\hat\nu)\rangle \enspace.
\label{eAmpere}
\end{equation}
Noting that $F(q)$ is the magnetic permeability, we have
\begin{equation}
F(q)={\langle A_\mu(q\hat\nu)\rangle\over A_\mu^{\rm ext}(q\hat\nu)}
={1\over 1-\langle A_\mu^{\rm ind}(q\hat\nu)\rangle/\langle 
A_\mu(q\hat\nu)\rangle}
\enspace.
\label{eF1}
\end{equation}
Combining this with Eqs.\,(\ref{eLondon}) and (\ref{eAmpere})
gives
\begin{equation}
  F(q)= {\alpha(q)Q^2\over 1+\alpha(q)Q^2}\enspace.
\label{eF2}
\end{equation}
Comparing with the results of Fig.\,\ref{f4}, we see that
for finite $\xi$ at $T<T_c$, we have $\lim_{q\to 0}\alpha(q)\sim\xi$.
Eq.\,(\ref{ealpha}) thus implies that the superconducting electron
density vanishes as $n_s\sim\xi^{-1}\sim|t|^{\nu}$.

The renormalized magnetic penetration length $\lambda$ is
determined by the pole of $F(q)$.  If one could ignore the
$q$ dependence of $\alpha(q)$, one would then conclude that
$\lambda^2=\alpha(q=0)$.  From this follows $\lambda\sim\sqrt\xi$
and $n_s\sim\lambda^{-2}$.  These are indeed the expectations
from mean field theory \cite{R11}, as well as the ``uncharged'' 
superconductor represented by the ordinary 3D XY model \cite{R12}
(given by the limit $\lambda_0\to\infty$).  They 
also hold in the present model, at low temperatures.

However, as $T\to T_c^-$, Eq.\,(\ref{eF2}) and Fig.\,\ref{f4}
imply that $\lim_{q\to 0}\alpha(q)\sim 1/q$.  This is
a consequence of the anomalous scaling dimension of the magnetic
field.  It is this singular dependence
of $\alpha(q)$ on $q$ that shifts
the pole of $F(q)$ so that $\lambda\sim\xi$ rather than $\sqrt\xi$,
in the ``charged'' superconductor critical region.  In this 
critical region the London relation $n_s\sim \lambda^{-2}$ no
longer holds \cite{R13}.

Note also that $F(q)$ determines the decay of
the magnetic field away from a test vortex line,
hence it determines the renormalized interaction between
vortex lines. The $q$ dependence of $\alpha(q)$ implies
that in the critical region, the interaction between 
two straight and parallel test vortex lines separated by distance $r$
will change from the $\ln r$ of mean field (MF) theory at $r<\sqrt\xi\sim
\lambda_{\rm MF}$, to
the faster decay of $1/r$ for $\sqrt\xi<r<\xi\sim\lambda$.

\begin{figure}
\epsfxsize=7.5truecm
\epsfbox[92 322 586 660]{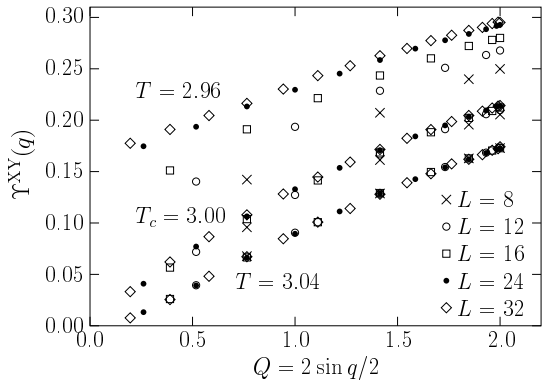}
\caption{$\Upsilon^{\rm XY}(q)$ vs.\ $Q$ for the ordinary
3D XY model. Data is for $L=8-32$ at
$T=2.96<T_c$, $T_c=3.0$ and $T=3.04>T_c$.  
Note the virtual absence of finite size effects for
$T\ge T_c$.
}
\label{f5}
\end{figure}

Finally, we note that the anomalous scaling of $F(q)$
also has some interesting consequences for the ordinary
3D XY model.  One can show that, within the mapping
of the 3D XY model to a gas of sterically interacting loops,
the helicity modulus of the XY model maps into a loop-loop
correlation function.  Identifying such loops as the vortex
lines of the LLS, which as $q\to 0$ (or $\lambda_0\to 0$)
become identical with magnetic flux, one concludes \cite{R14}
that the {\it wavevector dependent} helicity modulus \cite{R10,R15}
$\Upsilon^{\rm XY}(q)$ of the ordinary 3D XY model should be the
dual of $F(q)$.  We have carried out independent MC simulations of
the ordinary 3D XY model, in the Villain approximation,
calculating $\Upsilon^{\rm XY}(q)$ for
an ensemble with ``fluctuating twist'' boundary conditions (fbc) 
\cite{R15}.
We plot our results for $\Upsilon^{\rm XY}(q)$ vs.\ $Q$ in Fig.\,\ref{f5},
for $L=8-32$ and $T=2.96<T_c$, $T_c=3.0$ and $T=3.04>T_c$.
Note the striking similarity to Fig.\,\ref{f2}, only with the temperature
scale inverted. Finite size effects are negligible for $T\ge T_c$.
As $Q\to 0$, $\Upsilon^{\rm XY}$ approaches a constant for $T<T_c$,
$\Upsilon^{\rm XY}\sim Q^2$ for $T>T_c$, and $\Upsilon^{\rm XY}\sim Q$
exactly at $T_c$.  Thus the anomalous scaling of $F(q)$ at $T_c$
shows up as an anomalous scaling of $\Upsilon^{\rm XY}(q)$ at $T_c$
\cite{R16}.

To conclude, we have presented MC data that verifies the scaling
Ansatz of Eq.\,(\ref{eFscale}).  This Ansatz implies that there
is only a single diverging length scale in the problem, and that the
magnetic penetration length scales as $\lambda\sim\xi\sim|t|^{-\nu}$.
We find the value of $\nu=0.66\pm 0.03$ consistent with 
$\nu_{\rm XY}\approx 2/3$ of the
ordinary 3D XY model, and confirm the predicted anomalous scaling
dimension of magnetic field correlations at $T_c$.

We are indebted to I. F. Herbut and Z. Te\v{s}anovi\'{c} for
valuable discussions and a critical reading of the manuscript.
This work has been supported by U.S. DOE grant DE-FG02-89ER14017,
by Swedish Natural Science Research Council Contract No. E-EG 10376-305,
and by the resources of the Swedish High Performance Computing Center 
North (HPC2N).







\begin{thebibliography}{99}
\bibitem{R16}Our results for the ordinary 3D XY model show that,
    with fbc, the critical region in $q$ is smaller than
    was found for the LLS model.  Hence our largest size $L=32$
    is still too small for us to see a good scaling collapse at $T_c$,
    as in the corresponding case of Fig.\,\protect\ref{f3}.
\end{thebibliography}
\end{document}